\documentclass[journal]{IEEEtran}
\usepackage[utf8]{inputenc}
\usepackage{graphicx}
\usepackage{xcolor}
\usepackage{amsmath}
\usepackage{multirow}

\begin{document}

\title{GCN-ABFT: Low-Cost Online Error Checking for Graph Convolutional Networks}
\author{Christodoulos Peltekis and Giorgos Dimitrakopoulos
\thanks{This work was supported by a research grant of Siemens EDA to Democritus University of Thrace for ``HLS Research for Systems-on-Chip''.}
\thanks{%
Christodoulos Peltekis and Giorgos Dimitrakopoulos are with the Department
of Electrical and Computer Engineering, Democritus University of Thrace, Xanthi, Greece.\protect\\
E-mail: \{cpeltekis, dimitrak\}@ee.duth.gr}}

\maketitle

\begin{abstract}
Graph convolutional networks (GCNs) are popular for building machine-learning application for graph-structured data. 
This widespread adoption led to the development of specialized GCN hardware accelerators. In this work, we address a key architectural challenge for GCN accelerators: how to detect errors in GCN computations arising from random hardware faults with the least computation cost.  
Each GCN layer performs a graph convolution, mathematically equivalent to multiplying three matrices, computed through two separate matrix multiplications.
Existing Algorithm-based Fault Tolerance (ABFT) techniques can check the results of individual matrix multiplications. However, for a GCN layer, this check should be performed twice. To avoid this overhead, this work introduces GCN-ABFT that directly calculates a checksum for the entire three-matrix product within a single GCN layer, providing a cost-effective approach for error detection in GCN accelerators.
Experimental results demonstrate that GCN-ABFT reduces the number of operations needed for checksum computation by over 21\% on average for representative GCN applications. These savings are achieved without sacrificing fault-detection accuracy, as evidenced by the presented fault-injection analysis.

\end{abstract}

\begin{IEEEkeywords} Graph Convolution Networks, Algorithm-based Fault Tolerance, Energy Efficiency.
\end{IEEEkeywords}

\section{Introduction}
Machine learning (ML) has witnessed a surge in popularity in recent years.
Graph Neural Networks (GNNs) have emerged as a powerful tool for analyzing graph-structured data~\cite{recommend}.
GNNs excel at learning graph representations, where nodes correspond to objects and edges capture the relationships between them. 
This allows them to tackle tasks like graph classification, node classification, and link prediction~\cite{classification}. 
Each GNN layer updates node features by incorporating information from the features of neighboring nodes. By stacking multiple GNN layers, the model can effectively capture the influence of nodes further away in the graph~\cite{Kipf}.
In this work, we focus on Graph Convolutional Networks (GCNs)~\cite{Kipf} that
iteratively update the feature representations of all nodes in the graph
by multiplying the normalized adjacency matrix representing graph structure, the feature matrix of the preceding GCN layer, and a trainable weight matrix specific to the current layer.

The widespread adoption of GCNs has accentuated the need to accelerate them directly in hardware~\cite{gnn-accel, grow, sgcn, survey}. State-of-the-art GCN accelerators compute the three-matrix product needed per GCN layer using a two-phase process~\cite{gnn-accel, survey, fused, combination-first}: Aggregation phase involves a multiplying the normalized adjacency matrix with the feature matrix, while the combination phase refers to the multiplication of the features with the weights. These distinct phases exhibit contrasting dataflows and sparsity patterns, which has motivated the creation of diverse GCN accelerators~\cite{fused, combination-first, gcnax}. Combination-first computation exhibits the lowest arithmetic intensity and is preferred in recent GCN accelerators~\cite{combination-first}.

In addition to performance and energy efficiency, modern accelerators must also prioritize reliability~\cite{su2023testability}. This work addresses a critical architectural challenge for GCN accelerators: 
how to check a GCN layer's correctness in the presence of random hardware faults, while minimizing checking overhead.
Algorithm-Based Fault Tolerance (ABFT)~\cite{abft, abft-sa} offers a cost-efficient solution for detecting errors in matrix computations~\cite{abft-practical, convGuard}. It achieves this by comparing the checksum of the true output with a predicted one. ABFT has been successfully applied to verify two-matrix multiplications involving both dense and sparse data~\cite{braun2014abft, low-voltage-sa, kosaian2021arithmetic, aicas-abft}. 

In this work, we propose a low-cost adaptation of ABFT to the three-matrix product of a graph convolution, enabling checksum prediction for GCN layer as a whole.
The contributions of this work can be summarized as follows:
\begin{itemize}
\item
GCN-ABFT is a custom-designed ABFT method for GCN layers. Unlike traditional approaches that verify each matrix multiplication separately, GCN-ABFT calculates a predicted checksum for the entire three-matrix product of graph convolution in a single combined step.
\item
This fused checksum computation reduces the number of operations required for error checking compared to separate checks for the two individual matrix multiplications by 21\% on average for representative applications.
\item
Experimental results demonstrate that injecting single bit flips into arithmetic operations, such as matrix multiplications and checksum computations, at random times improves slightly the fault detection accuracy of GCN-ABFT compared to baseline ABFT. This improvement is attributed to the reduced check state required by GCN-ABFT, which leads to fewer false error detections.

\end{itemize}

\section{Applying ABFT to GCNs}
Since a GCN layer involves two separate matrix multiplications, traditional error checking using ABFT would require calculating checksums twice: once for each multiplication step.

\subsection{Computations involved in a GCN Layer}
In each GCN layer, each node first collects the features of all its neighboring nodes and reduce them to a new feature vector using averaging or max/min operators~\cite{gnn-accel, Kipf}. This aggregation operation can be expressed over the entire graph at the $k$th GCN layer as a matrix multiplication $\tilde{H} = S\, H^{k-1}$. $H^{k-1}$ denote the output features of all nodes of the previous $(k-1)$th GCN layer and $S$ represents the normalized adjacency matrix: $S = D^{1/2}\tilde{A} D^{1/2}$, with $\tilde{A} = A+I$ and $D$ being the degree matrix of the graph adjacency matrix $A$. 

The aggregated features of all nodes $\tilde{H}$ are then linearly transformed using the weight matrix $W^k$ of the $k$th layer, and passed through a non-linear activation function $\sigma$:
\begin{equation}
H^k = \sigma \left(\tilde{H} W^k \right) = \sigma \left(S H^{k-1} W^{k} \right)
\label{e:main}
\end{equation}
To simplify notation, we henceforth remove the layer indices from matrices $H$ and $W$ and denote the targeted three-matrix multiplication of a GCN layer as $H_{out} = S\, H\, W$.

\subsection{Error checking GCN layers with baseline ABFT}
The straight-forward choice to apply error checking on graph convolution of Eq.~\eqref{e:main} (before the application of the activation function), is to apply ABFT separately on the two phases of the computation.

Assuming a combination-first computation, which is the preferred dataflow in recent GCN accelerators~\cite{combination-first} and requires the less operations in many  applications, we should first compute the $X = H\, W$ and then compute the output feature matrix as $H_{out} = S\, X$. 

Adhering to the ABFT methodology~\cite{abft}, for validating the first matrix multiplication $X = H\, W$, it is essential to compare the \emph{actual checksum} of all elements of the output matrix $X$ with a \emph{predicted checksum} derived independently by the elements of $H$ and $W$. The final checksum of $X$ can be computed as $e^T X e$ with $e^T = [1, 1, \ldots 1]$. Expanding the checksum equation we get that $e^T\! X e\!=\! e^T\! H W e \!=\! (e^T\! H)(W e) \!=\! h_c w_r$. To compute this checksum, it suffices to enhance $H$ with an extra row that represents the per-column checksum vector of $H$, i.e., $h_{c} = e^T H$, and matrix $W$ with an extra column that includes the corresponding per-row checksum of $W$, i.e., $w_r = W\, e$. 
Since the weights are known beforehand $w_r$ can be computed offline or during weight loading to the GCN accelerator. On the contrary, the per-column checksum $h_c$ of $H$ can be computed only \emph{online} (except only for the first GCN layer), since the input matrix $H$ of a GCN layer is the output matrix of the previous layer. 

\begin{figure}
\centering
\includegraphics[width=0.65\columnwidth]{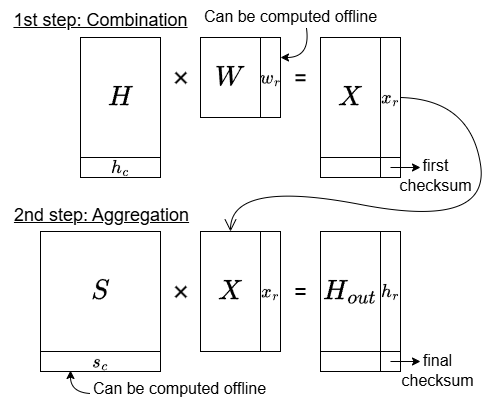}
\caption{ABFT applied separately on the two phases of graph convolution operation.}
\label{f:separate}
\end{figure}

Performing matrix multiplication with the enhanced matrices $H$ and $W$ leads to
\begin{equation}
\begin{bmatrix}    
H \\ h_c
\end{bmatrix}
\begin{bmatrix}    
W & w_r
\end{bmatrix}=
\begin{bmatrix}    
\mathbf{H W} &  H w_r \\
h_c W &  \mathbf{h_c w_r} 
\end{bmatrix}
\label{e:abft-first}
\end{equation}
Since in~\eqref{e:abft-first} matrices $H$ and $W$ are enhanced with their extra check state, the final checksum $h_{c} w_{r}$ of $X$ is naturally computed at the lower right end of the resulting matrix.
This operation is also graphically depicted in Fig.~\ref{f:separate}. To identify any erroneous result, the actual checksum accumulated online should be compared with the predicted checksum $h_c w_r$.

Completing the computation of the output features $H_{out}$ for a GCN layer requires multiplying the normalized adjacency matrix $S$ with the output of the previous step, i.e., $H_{out} = S\, X$. To check this multiplication with ABFT, we should enhance $S$ with the per-column checksum vector $s_c = e^T\! S$ and $X$ with the per-row checksum $x_{r} = X e = H W e = H w_r$. 
For static graphs $s_{c}$ can be computed offline and reused.
Also, $x_{r} = H w_r$ is already computed online at the upper right part of the output of~\eqref{e:abft-first}.
Multiplying the two enhanced matrices is performed as follows (shown also in Fig.~\ref{f:separate}):
\begin{equation}
\begin{bmatrix}    
S \\ s_c
\end{bmatrix}
\begin{bmatrix}    
X & x_r
\end{bmatrix}=
\begin{bmatrix}    
\mathbf{S X} &  S x_r \\
s_c X&  \mathbf{s_c x_r} 
\end{bmatrix}
\label{e:abft-second}
\end{equation}

Fault detection for this GCN layer occurs after finding a discrepancy between the predicted checksum $s_c x_r$ and the true checksum of $H_{out}$ computed online.

\section{GCN-layer specific ABFT}
\label{s:fused-abft}

Graph convolution is computed using two matrix multiplication steps. However, we are not obliged to apply ABFT at each step separately. Instead, we design a new error checker that computes the checksum of the three-matrix multiplication $H_{out} = S\, H\, W$ for a GCN layer in a \textit{fused} manner. 
This consolidation of checksum computation removes unnecessary computations found in the split checking process, thus reducing significantly the overall number of operations required for applying ABFT in GCNs.

The checksum $H^{C}_{out}$ of the output of a GCN layer $H_{out}$ is equal to
\begin{equation}
H^{C}_{out} \!=\! e^T\! H_{out} e \!=\! e^T\! (S H W) e \!=\! (e^T\! S) H (W e) \!=\! s_{c} H w_{r}
\label{e:checksum-fused}
\end{equation}
$s_c =(e^T\! S) $ and $w_r = W e$ correspond to the per-column checksum vector of the normalized adjacency matrix $S$ and the per-row checksum vector of the weight matrix $W$, respectively. 

\begin{figure}
\centering
\includegraphics[width=0.65\columnwidth]{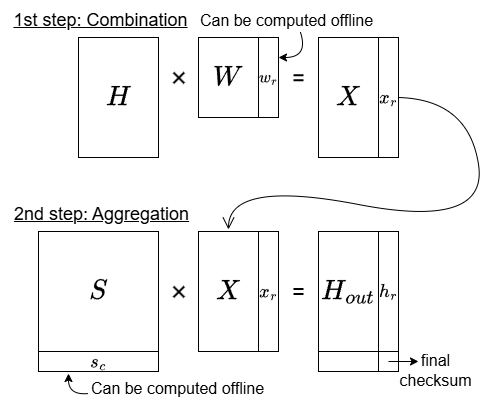}
\caption{GCN-ABFT applied in the two phases of graph convolution. Fusing checksum computation removes the need to enhance matrix $H$ with additional check state.}
\label{f:fused-checksum}
\end{figure}

In GCN-ABFT, based on the fused checksum~\eqref{e:checksum-fused} and assuming the same dataflow as the baseline approach, we perform checksum computation as follows: $H$ is multiplied \emph{as is} and \emph{without any check state} with the enhanced 
weight matrix $W$ that includes also the per-row checksum vector $w_r$. This operation is performed as follows:
\begin{equation}
H\, \begin{bmatrix}
W & w_r    
\end{bmatrix}= 
\begin{bmatrix}
\mathbf{H\,W} & \mathbf{H\, w_r}     
\end{bmatrix}
\label{e:fused-step-1}
\end{equation}
The resulting matrix $[H W \quad  H w_r]$ includes both the actual result of $X \!=\! H\, W$ and its check vector $H\, w_r$ that corresponds to the right part of~\eqref{e:checksum-fused}. In fact, the computed check vector $H\, w_r$ corresponds to the per-row checksum vector $x_r$ of the intermediate output $X$ since $x_r = X e  = H W e = H (W e) = H w_r$. This feature is graphically depicted in Fig.~\ref{f:fused-checksum}.

In the next step, the normalized adjacency matrix $S$ is multiplied with the output of the first step. To enable ABFT, we enhance $S$ with an extra row that corresponds to the per-column checksum of its elements $s_c$. The enhanced version of $[X\quad x_r] = [HW\quad Hw_r]$ is already available by~\eqref{e:fused-step-1}. The multiplication of the enhanced matrices $S$ and $X$ is described as follows:
\begin{equation}
\begin{bmatrix}
S \\ s_c    
\end{bmatrix} 
\begin{bmatrix}
H\,W & H\, w_r  
\end{bmatrix} =
\begin{bmatrix}
\mathbf{S\,H\,W} & S\, H\, w_r \\
s_c\,H\,W & \mathbf{s_c\, H\, w_r}
\end{bmatrix}
\label{e:fused-step-2}
\end{equation}

This matrix multiplications computes at its lower right part the fused checksum $s_c H w_r$ of Eq.~\eqref{e:checksum-fused}.
Therefore, the final checksum is computed without relying on any intermediate per-column checksum for matrix $H$ of each GCN layer.
The removal of this check state for $H$ removes redundant computations and makes the checker less vulnerable to faults that may affect its check state. Also, GCN-ABFT relies only on the per-column and per-row checksum vector of statically defined matrices $S$ and $W$. Therefore, it is easier for GCN-ABFT to compute such vectors offline and reuse them for multiple GCN inferences.

Overall, GCN-ABFT is a generic approach that can be applied to aggregation-first dataflows as well, since the computation of the fused checksum given by~\eqref{e:checksum-fused} holds independent of the order of computations and the structure or sparsity of the involved matrices.
The fused checksum computation of GCN-ABFT may possibly be adapted to other classes of GNNs that their operation can be expressed as a three-matrix product~\cite{ethz-gnn}.

The adoption of GCN-ABFT introduces two minor trade-offs. 
First, GCN-ABFT may miss a certain fault case that can be detected by baseline ABFT, which checks each matrix multiplication separately. A fault can go undetected by GCN-ABFT when the normalized adjacency matrix $S$ contains a column filled entirely with zeros. In this scenario, any fault affecting the first multiplication $H\, W$ is nullified at the output $S (H\,W)$, preventing its detection. Baseline ABFT, on the other hand, would have detected the fault in the first matrix multiplication. While this case is theoretically possible, it cannot occur in valid GCN scenarios as no column in $S$ can be entirely filled with zeros.

The second tradeoff refers to the timeline of error detection in GCN-ABFT.
If an error occurs in the first matrix multiplication step $X\!=\! H W$, the baseline ABFT can report the error directly and avoid performing the second matrix multiplication $H_{out}\!=\! S X$. With GCN-ABFT this is not possible and error detection can be reported only \emph{at the end of each GCN layer}, after both matrix multiplications per layer are completed.

The fixed-delay reaction of GCN-ABFT (once per GCN layer) matters only in the rare event of an actual error occurrence. In the vast majority of cases no error happens during GCN execution and the energy spent for validating the output's correctness is redundant and should be minimized as much as possible. Thus, the savings offered by GCN-ABFT are truly important, since energy is saved when it matters most and that is during error-free operation.
Additionally, when the error occurs in the second matrix multiplication of graph convolution both baseline ABFT and GCN-ABFT detect the error at the same time and always within the time bounds of a GCN layer.

\section{Evaluation}
\label{s:evaluation}
The goal of the experimental results is to compare the performance of GCN-ABFT relative to baseline ABFT both in terms of fault-detection accuracy and the number of operations required to perform checking. This quantitative analysis was performed on four well-known GCN applications of varying complexity used for node classification: Cora, Citeseer, PubMed and Nell~\cite{graphlearning}.

\subsection{Experimental Setup}

\begin{table*}[t]
    \setlength{\tabcolsep}{4pt}
    \renewcommand{\arraystretch}{1}
    \centering
    \caption{Fault detection accuracy for representative GCN applications for a single injected fault using several error bounds}
    \begin{tabular}{c||c|c||r||c|c|c|c|c|c|c|c}
    \hline
          \multirow{2}{4em}{\centering GCN}
          & \multirow{2}{4em}{\centering Critical Faults} & 
          \multirow{2}{5em}{\centering Avg. Nodes Affected} & ~ & \multicolumn{2}{|c|}{$10^{-4}$} & \multicolumn{2}{|c|}{$10^{-5}$} & \multicolumn{2}{|c|}{$10^{-6}$} & \multicolumn{2}{|c}{$10^{-7}$} \\ \cline{5-12}
        ~ & ~ & ~ & ~ & Split & GCN-ABFT & Split & GCN-ABFT & Split & GCN-ABFT & Split & GCN-ABFT \\ \hline
        \multirow{3}{*}{Cora} & \multirow{3}{*}{96.92\%} & \multirow{3}{*}{68.61\%} & \textbf{Detected}  & 95.46\% & 96.42\% & 95.52\% & 96.06\% & 95.80\% & 96.66\% & 95.80\% & 96.66\% \\
                              &         &         & \textbf{False Pos} & 3.84\%  & 3.14\%  & 4.14\%  & 3.18\%  & 4.20\%  & 3.34\% & 4.20\%  & 3.34\% \\
                              &         &         & \textbf{Silent}    & 0.70\%  & 0.80\%  & 0.34\%  & 0.40\%  & 0.00\%  & 0.00\% & 0.00\%  & 0.00\% \\ \hline
    \multirow{3}{*}{Citeseer} & \multirow{3}{*}{92.60\%} & \multirow{3}{*}{33.70\%} & \textbf{Detected}  & 93.64\% & 94.42\% & 95.16\% & 96.72\% & 95.44\% & 97.06\% & 95.44\% & 97.06\% \\
                              &         &         & \textbf{False Pos} & 3.62\%  & 2.42\%  & 4.48\%  & 2.86\%  & 4.56\%  & 2.94\% & 4.56\%  & 2.94\% \\
                              &         &         & \textbf{Silent}    & 2.74\%  & 3.16\%  & 0.36\%  & 0.42\%  & 0.00\%  & 0.00\% & 0.00\%  & 0.00\% \\ \hline
      \multirow{3}{*}{PubMed} & \multirow{3}{*}{96.10\%} & \multirow{3}{*}{28.32\%} & \textbf{Detected}  & 94.32\% & 94.96\% & 94.72\% & 95.86\% & 96.14\% & 97.02\% & 96.38\% & 97.42\% \\
                              &         &         & \textbf{False Pos} & 3.06\%  & 2.54\%  & 3.08\%  & 2.56\%  & 3.54\%  & 2.56\% & 3.62\%  & 2.58\% \\
                              &         &         & \textbf{Silent}    & 2.62\%  & 2.50\%  & 2.20\%  & 1.58\%  & 0.32\%  & 0.42\% & 0.00\%  & 0.00\% \\ \hline
        \multirow{3}{*}{Nell} & \multirow{3}{*}{95.32\%} & \multirow{3}{*}{61.06\%} & \textbf{Detected}  & 94.88\% & 95.38\% & 95.40\% & 96.00\% & 96.84\% & 97.62\% & 96.90\% & 97.82\% \\
                              &         &         & \textbf{False Pos} & 2.58\%  & 1.78\%  & 2.64\%  & 1.92\%  & 3.02\%  & 2.14\% & 3.10\%  & 2.18\% \\
                              &         &         & \textbf{Silent}    & 2.54\%  & 2.84\%  & 1.96\%  & 2.08\%  & 0.14\%  & 0.24\% & 0.00\%  & 0.00\% \\ \hline
        
    \end{tabular}
    \label{t:detection-rate}
\end{table*}

To evaluate the quality of error detection for both approaches, we introduce random single-bit flips~\cite{bit-flips} into the results of arithmetic operations~\cite{abft-practical, kosaian2021arithmetic} within matrix multiplication (multiply and add) or checksum accumulation, at randomly selected time points per GCN layer. The affected arithmetic operations for matrix multiplications involve single-precision floats, while checksum accumulation uses double-precision floats.
All bits of every arithmetic operation output can be flipped with equal probability. 

A fault can occur at any time point during the execution of a GCN layer.
Consequently, faults are more likely to occur during the matrix multiplication step that lasts longer (involving larger matrices). Similarly, due to the higher number of multiply-add operations in matrix multiplication compared to checksum accumulation, faults are more likely to affect multiply-add operations.

We assume memory is protected by error detection~\cite{memory-protection} and thus input data fetched are fault-free. However, arithmetic faults on index increments that track non-zero elements in sparse matrices $S$ and $H$, which are stored in CSR format~\cite{fused}, can indirectly corrupt memory accesses. This can lead to incorrect data fetching that ABFT cannot detect. As a result, we did not inject such faults. In any case, address generation logic is relatively small compared to wide floating-point datapaths, making bit flips in this area infrequent.

After fault injection the observed behavior, at the end of a GCN layer, may fall into one of three categories:
\begin{enumerate}
\item 
Detected: A faulty output was computed and ABFT detected it. 
\item 
False positive: 
A fault injected into the checksum accumulation led ABFT to incorrectly identify a correct output result as erroneous.
\item 
Silent: Error remains undetected by ABFT. This case is a result of floating point rounding errors that does not allow to distinguish the effect of an injected fault from a correct result.
\end{enumerate}
To prevent silent faults during our fault-injection campaigns, we need to establish a threshold~\cite{braun2014abft} that differentiates between silent faults and other fault categories. This threshold, determined experimentally for specific GCN applications and datasets, considers error bounds ranging from $10^{-4}$ to $10^{-7}$. We found that setting the threshold above $10^{-7}$ eliminates silent faults in all GCNs.

False negative faults require a fault injected to matrix multiplication and checksum accumulation to cancel each other thus causing ABFT to fail to identify an actual fault in the output. Such cases were not observed in our experiments.

\subsection{Error detection quality}

Table~\ref{t:detection-rate} summarizes the categorization of the injected faults derived from 5000 independent fault-injection campaigns. Using more fault-injection campaigns do not change the observed behavior.
In this setup, a single fault was injected in each application. Which GCN layer actually experiences the fault is proportional to its execution time.
Fault injection results in 71.1\% of all possible bit flips in multiply-add outcomes and 55.8\% in the checksum accumulator, on average.

Columns 2--3 quantify how much critical were the injected faults for each GCN applications and Columns 5--12 report the fault detection accuracy of both ABFT variants.
A fault is critical if it changes the final classification of at least one node of the graph. For instance, as shown in Column 2 of Table~\ref{t:detection-rate}, in Cora, 96.92\% of the injected faults triggered the misclasification of at least one node of the graph. Additionally, in Column 3 of Table~\ref{t:detection-rate},
we report the average number of nodes of each graph that are critically affected by each injected fault.
In this way, we quantify the spread of the effect of the fault in all nodes of the graph. For instance, in Citeseer, each fault causes the misclassification of 33.7\% of the nodes, on average.

Fault criticality at the application level does not change the fault detection properties of ABFT. For instance, a fault may be non-critical at the application level and still be considered as detected since it was identified as ruining the result of matrix multiplication at the checksum level. 

Both techniques achieve high fault detection accuracy, exceeding 93\% in all cases.
Silent faults disappear in all applications, when assuming that a fault is detected if the absolute difference between the predicted and output checksum exceeds $10^{-7}$. In all cases, GCN-ABFT exhibits fewer false positives because it calculates the checksum using Equation~\eqref{e:checksum-fused} and doesn't require a check state for matrix $H$, unlike the baseline ABFT approach (Eq.~\eqref{e:abft-first}).

We performed additional experiments by injecting more than one single-bit arithmetic faults per GCN application in randomly selected time points. In such cases, fault detection for baseline ABFT and GCN-ABFT reaches 100\% offering an almost indistinguishable error detection quality.

\begin{table}[t]
\setlength{\tabcolsep}{4pt}
\renewcommand{\arraystretch}{1}
\centering
\caption{Millions of arithmetic operations needed for executing and validating representative GCN applications}
\begin{tabular}{c|c||c|c||c|c||c|c}
\hline
         &  True  & \multicolumn{2}{c||}{Split} & \multicolumn{2}{c||}{GCN-ABFT} & \multicolumn{2}{c}{Savings} \\ \cline{3-8}
GCN      &  Out   & Check  & Total & Check & Total  & Check   & Total \\ \hline
Cora     & 2.8 &     0.55  &  3.35   & 0.44 & 3.24  &  20.0\% & 3.3\% \\
Citeseer & 4.6 &     0.80  &  5.40   & 0.60 & 5.20  &  25.0\% & 3.7\% \\ 
Pubmed   & 37.6 &    4.60  & 42.20   & 4.04 & 41.64 &  12.2\% & 1.3\% \\
Neil     & 1745.9 & 84.30  & 1830.20 & 59.9 & 1805.8& 28.9\%  & 1.3\% \\ \hline
\end{tabular}
\label{t:number-of-ops}
\end{table}

\subsection{Computation cost}
Computing and validating a GCN layer using the GCN-ABFT's checksum computed via Eqs.~\eqref{e:fused-step-1} and~\eqref{e:fused-step-2} requires \emph{less operations in overall} than the split-checksum approach implemented using Eqs.~\eqref{e:abft-first} and~\eqref{e:abft-second}.

To quantify the efficiency gains of GCN-ABFT compared to baseline ABFT (which checks each matrix multiplication separately), we measured the total number of operations required in each case including:
(a) Operations essential for computing the actual GCN layer output.
(b) Operations needed to compute the actual checksum (sum of all output elements) and the predicted checksum at each check step. 
Table~\ref{t:number-of-ops} details the number of operations per category (Multiplications and additions are counted equally). 
GCN-ABFT demonstrates savings in checksum computation that range between 12\% and 28\% and translate to 1.3\%--3.7\% savings in overall computation cost.
This efficiency improvement stems from the fused-approach used by GCN-ABFT that computes one check per GCN layer:
(a) actual checksum computations are halved because GCN-ABFT only verifies the final GCN layer output, requiring the sum of elements only from the second matrix multiplication; (b) the reduced check state needed by Eq.~\eqref{e:fused-step-1} compared to Eq.~\eqref{e:abft-first} translates to fewer operations for calculating the predicted checksum.

\begin{figure}
\centering
\includegraphics[width=0.68\columnwidth]{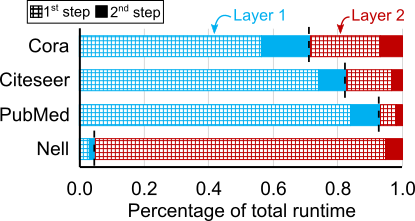}
\caption{How the execution time is split across the first and the second matrix multiplication step of each GCN layer for both layers of the examined GCN applications.}
\label{f:notification-gap}
\end{figure}

\subsection{Error-detection latency gap between GCN-ABFT and baseline ABFT}

GCN-ABFT reliably detects errors at the end of each GCN layer, regardless of the error's timing within the two multiplication steps. While baseline ABFT could potentially detect first-step errors earlier, the domination of the runtime of
first multiplication phase (combination) in each GCN layer, as shown in Fig.~\ref{f:notification-gap}, makes this advantage negligible.

Fig.~\ref{f:notification-gap} shows the time spent at each multiplication step of a GCN layer normalized to the total runtime of the specific layer for all examined applications. The applications examined are two-layer GCNs. The textured regions represent the runtime of the first multiplication step in both layers of the examined GCNs. The uniformly colored regions represent the second multiplication step in each case. In all cases, the first multiplication step of both layers dominates the overall runtime.  
For instance, for PubMed, the first multiplication step of both layers are responsible for more than the 90\% of the runtime, while, for Nell the first multiplication step of both layers is 95\% of the overall execution time. 
Therefore, while baseline ABFT can signal errors after the first multiplication, GCN-ABFT's delay in waiting for the second multiplication as well, is negligible per layer, making its fixed-delay response a minor issue.

\section{Conclusions}
GCN-ABFT demonstrates that, for a GCN layer involving two consecutive matrix multiplications, it is more efficient to compute the checksum of the final result directly rather than verifying each step individually. We validated this approach across various GCN applications, significantly reducing the number of operations required for online verification of GCN inferences without compromising fault detection capabilities. Furthermore, evaluation indicates that the fixed-delay response of GCN-ABFT has minimal impact on practical applications. Since the initial multiplication step within each GCN layer consumes a substantial portion of the runtime, the difference in response delay between GCN-ABFT and the baseline two-step ABFT becomes negligible.

\bibliographystyle{IEEEtran}
\bibliography{refs}

\end{document}